\documentclass[twocolumn]{aastex63}

\received{Month Date, 2021}
\revised{Month Date, 2022}
\accepted{Month Date, 2022}
\submitjournal{ApJ}

\shorttitle{{\textsc{21cmVAE}}}
\shortauthors{Bye et al.}

\graphicspath{{./}}

\begin{document}

\title{{\textsc{21cmVAE}}: A Very Accurate Emulator of the 21-cm Global Signal}

\correspondingauthor{Christian H. Bye}
\email{chbye@berkeley.edu}

\author[0000-0002-7971-3390]{Christian H. Bye}
\affiliation{Department of Astronomy, University of California, Berkeley, CA 94720, USA}
\affiliation{Department of Physics, McGill University, Montréal, QC H3A 2T8, Canada}

\author[0000-0001-8132-8056]{Stephen K. N. Portillo}
\affiliation{DIRAC Institute, Department of Astronomy, University of Washington, 3910 15th Ave. NE, Seattle, WA 98195, USA}

\author[0000-0002-1369-633X]{Anastasia Fialkov}
\affiliation{Institute of Astronomy, University of Cambridge, Madingley Road, Cambridge CB3 0HA, United Kingdom}
\affiliation{Kavli Institute for Cosmology, Madingley Road, Cambridge CB3 0HA, UK}

\begin{abstract}
Considerable observational efforts are being dedicated to measuring the sky-averaged (global) 21-cm signal of neutral hydrogen from  Cosmic Dawn and the Epoch of Reionization. Deriving observational constraints on the astrophysics of this era requires modeling tools that can quickly and accurately generate theoretical signals across the wide astrophysical parameter space. For this purpose artificial neural networks were used to create the only two existing global signal emulators, {\textsc{21cmGEM}} and {\textsc{globalemu}}. In this paper we introduce {\textsc{21cmVAE}}, a neural network-based global signal emulator, trained on the same dataset of $\sim 30,000$ global signals as the other two emulators, but with a more direct prediction algorithm that prioritizes accuracy and simplicity. Using neural networks, we compute derivatives of the signals with respect to the astrophysical parameters and establish the most important astrophysical processes that drive the global 21-cm signal at different epochs. {\textsc{21cmVAE}} has a relative rms error of only 0.34\%---equivalently 0.54 mK---on average, which is a significant improvement compared to the existing emulators, and a run time of 0.04 seconds per parameter set. The emulator, the code, and the processed datasets are publicly available at \url{https://github.com/christianhbye/21cmVAE} and through \url{https://zenodo.org/record/5904939}. 
\end{abstract}

\keywords{early universe --- 
cosmology --- astronomy software}

\section{Introduction}
\label{sec:intro}
The 21-cm line of neutral hydrogen is one of the most promising probes of Cosmic Dawn and the Epoch of Reionization (EoR). Emitted by neutral hydrogen at redshifts $z>6$, the signal is redshifted to frequencies below 200 MHz, and can therefore be observed by radio telescopes \citep{2000ApJ...528..597T}. The global 21-cm signal, obtained by averaging the spectrum across all sky, traces the cosmology and astrophysics of the high-redshift universe. The intensity of the signal is observed in contrast to the background radiation, which is normally assumed to be the CMB, and is quantified by the differential brightness temperature. This temperature depends on the excitation temperature of the 21-cm transition (the spin temperature), the temperature of the background radiation, and the abundance of neutral hydrogen. During Cosmic Dawn and the Epoch of Reionization, processes including the Wouthuysen-Field effect \citep[][]{1952AJ.....57R..31W, 1959ApJ...129..536F}, X-ray heating \citep{1997ApJ...475..429M}, and reionization of neutral hydrogen affect the differential brightness temperature and leave characteristic features in the global signal \citep{1999A&A...345..380S}. The Wouthuysen-Field effect describes how the spin states of neutral hydrogen are mixed through absorption and re-emission of Ly$\alpha$ photons, thus coupling the spin temperature to the kinetic temperature of the hydrogen gas. This effect was dominant in the beginning of the Cosmic Dawn when the gas temperature was cooler than the background radiation temperature, hence making the differential brightness temperature negative. Later, the first X-ray sources heated the intergalactic medium (IGM), making the gas go from absorption to emission against the background. Finally, the abundance of neutral hydrogen decreased due to reionization of the gas and the signal vanished. We refer to \citet{2006PhR...433..181F}, \citet{2012RPPh...75h6901P}, and \citet{2016PhR...645....1B} for in-depth reviews of these physical processes.

There are several ongoing efforts to detect the global signal with one reported detection \citep{2018Natur.555...67B}, made by the Experiment to Detect the Global EoR Signature \citep[EDGES,][]{2010Natur.468..796B, 2017ApJ...847...64M, 2018ApJ...863...11M, 2019ApJ...875...67M}. Other experiments include PRIZM \citep[Probing Radio Intensity at high-Z from Marion,][]{2019JAI.....850004P}, MIST (Mapper of the IGM Spin Temperature, \url{http://www.physics.mcgill.ca/mist/}), SARAS \citep[Shaped Antenna measurement of the background RAdio Spectrum,][]{2021arXiv211206778S}, REACH \citep[Radio Experiment for the Analysis of Cosmic Hydrogen,][]{deLeraAcedo2019, 2021arXiv210910098C}, and DAPPER \citep[Dark Ages Polarimeter PathfindER,][]{2019AAS...23421202B}. The EDGES detection showed a deep and narrow absorption profile, which cannot be explained by standard cosmological models. In particular, the lower 99\% confidence bound on the best-fit amplitude was 0.3 K, which is approximately 50\% greater than the largest predicted amplitude \citep{2018Natur.555...67B}. 

With the surprising first results and awaited new measurements, it is important to model the range of possible 21-cm signals from the early universe and explore the associated astrophysical parameter space. To this end, a flexible method is required to realize the global signal efficiently, allowing parameters to be constrained from measurements with the use of sampling techniques such as MCMC \citep{2020PASP..132f2001L}. Simulations of the 21-cm signal \citep[e.g.][]{2012Natur.487...70V, 2014Natur.506..197F} take a few hours to run and, thus, cannot be directly employed in a parameter estimation pipeline. Instead, emulators trained on the results of these simulations are becoming a popular tool for fast model generation. Currently available software include {\textsc{emupy}} \citep{2017ApJ...848...23K}, which emulates the 21-cm power spectrum; {\textsc{21cmGAN}} \citep{2020MNRAS.493.5913L}, which generates tomographic samples of the 21-cm brightness temperature; and the two global signal emulators {\textsc{21cmGEM}} \citep{2020MNRAS.495.4845C} and {\textsc{globalemu}} \citep{2021MNRAS.508.2923B}. Given seven astrophysical parameters, {\textsc{21cmGEM}} calculates five auxiliary parameters and uses a series of neural networks, a bagged tree classifier, and principal component analysis to emulate the global signal over redshifts $z=5-50$. It is fast and accurate: the emulator has a running time of 0.16 seconds per parameter set and an average relative error of 1.59\%. The emulator has already been used to constrain the astrophysical parameters using the data of the EDGES High Band experiment \citep{2019ApJ...875...67M}. {\textsc{globalemu}} emulates the global signal from the same parameters as {\textsc{21cmGEM}}, using just a single neural network. The simpler prediction algorithm emulates the global signals significantly faster, with a run time of only 1.3 ms, and with a smaller average relative error of 1.12\%.

In this paper we present {\textsc{21cmVAE}}, a new emulator of the global signal from the same seven astrophysical parameters used in \textsc{21cmGEM} and \textsc{globalemu}: the star formation efficiency ($f_*$), the minimum circular velocity of star-forming halos ($V_c$), the X-ray radiation efficiency ($f_X$), the optical depth ($\tau$) of the Cosmic Microwave Background radiation (CMB), the power-law slope ($\alpha$) and low energy cutoff ($\nu_{\small \textrm{min}}$) of the X-ray spectral energy distribution (SED), and the mean free path of ionizing photons ($R_{\small \textrm{mfp}}$). The emulator uses the same dataset as {\textsc{21cmGEM}} and \textsc{globalemu}, with global signals simulated from the parameters using the method described in e.g. \citet{2020MNRAS.495.4845C}. We refer to \citet{2020MNRAS.495.4845C} for more details on the modeling.

The objective of this work is to use artificial neural networks to learn relationships between parameters and signals without enforcing any physical models. Compared to {\textsc{21cmGEM}} we aim to use a simpler prediction algorithm that does not calculate auxiliary parameters or depend on the input parameters. However, unlike {\textsc{globalemu}}, we do not aim to make the fastest or simplest model that meets a target performance---instead we explore a wider range of models and prioritize very accurate predictions. As a result, our emulator uses only one neural network which predicts global signals given the seven astrophysical parameters. This allows {\textsc{21cmVAE}} to be flexible and predict a wide range of signals. The emulator is written in Python, using the machine learning libraries TensorFlow \citep{tensorflow2015-whitepaper} and Keras \citep{chollet2015keras}, making it easy to modify or retrain. {\textsc{21cmVAE}} is available on \href{https://github.com/christianhbye/21cmVAE}{GitHub}, where the dataset and documentation also can be found.

This paper has six sections. After the introduction, section \ref{sec:methods} discusses how we designed and optimized our models and the performance and speed of the emulator is presented in section \ref{sec:results}. In section \ref{sec:impact}, we evaluate the impact of the astrophysical parameters on the global signal and in section \ref{sec:tsne}, we use the emulator to interpret an intermediate (latent) representation of the signals. We conclude in section \ref{sec:conclusions}. 
\section{Methods}
\label{sec:methods}
This section assumes familiarity with basic principles and terminology of neural networks, we refer to \citet{Goodfellow-et-al-2016} for a review of this topic.

\subsection{Architecture} \label{subsec:architecture}
The emulator has a 7-dimensional input layer (for 7 astrophysical parameters) and a 451-dimensional output layer, since it outputs realizations of global signals at 451 redshifts. It also has four hidden layers---of 288, 352, 288, and 224 dimensions, respectively---which all are fully connected and use the activation function ReLU \citep{4082265}. We also trained an emulator which utilized an autoencoder to create a low-dimensional representation of the global signals; this emulator appeared to be both less accurate and slower, but is described in Appendix \ref{appendix:21cmAE}. We earlier tried to use a variational autoencoder \citep{2013arXiv1312.6114K}, which uses a regularized latent space that we believed would preserve data structures better than the latent space of a vanilla autoencoder and thus work better in the emulator. However, we found that this regularization came at the cost of increased reconstruction error of the autoencoder and that the emulator error was larger with a variational autoencoder than a standard autoencoder.

We use the root mean squared (rms) error as a fraction of signal amplitude to evaluate the performance of the emulator. This is defined for each signal by:
\begin{equation}
\label{eq:fom}
    \textrm{Error} = \frac{\sqrt{{\langle (T(\nu) - \hat{T}(\nu))^2 \rangle}}}{\max(|T(\nu)|)}.
\end{equation}

\noindent Here, $\hat{T}(\nu)$ denotes the frequency-dependent signal predicted by the emulator, whereas $T(\nu)$ represents the simulated signal from the dataset. The figure of merit is the same as used by \textsc{21cmGEM}, making the results directly comparable. 

\subsection{Dataset} \label{subsec:dataset}
\begin{figure}
    \centering
    \includegraphics[width=\linewidth]{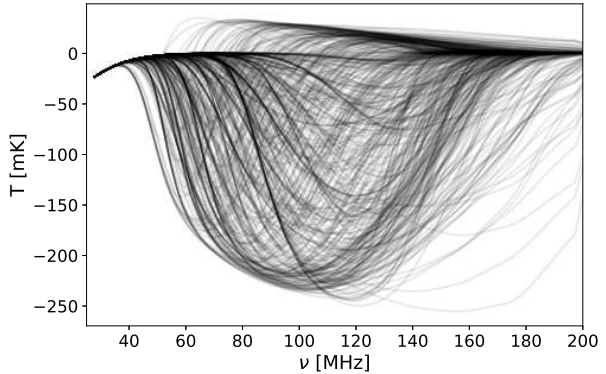}
    \caption{500 global signal models randomly drawn from the training set.}
    \label{fig:data_density}
\end{figure}

We use the publicly available dataset created for {\textsc{21cmGEM}} of $29,641$ signals, including a training set with $27,455$ signals and a test set with $2,186$ signals. We restrict the datasets to only include signals in the parameter ranges given by \citet{2020MNRAS.495.4845C} (the ranges for the test set in brackets): $f_* = 0.0001-0.50\; (0.0003-0.50)$, $V_c = 4.2-100\; (4.2-76.5)$ km~$\textrm{s}^{-1}$, $f_X = 0-1000\; (0-10)$, $\tau = 0.04-0.2\; (0.055-0.1)$, $\alpha = 1.0-1.5\; (1.0-1.5)$, $ \nu_{\small \textrm{min}} = 0.1-3.0\; (0.1-3.0)$ keV, $R_{\textrm{\small mfp}} = 10-50\; (10-50)$ Mpc. Figure \ref{fig:data_density} shows a representative sample of the global signal models in the training set.

Approximately $10\%$ of the training set is used as a validation set. This results in a total of $28,996$ global signals split into a training set with 24,562 signals, a validation set with $2,730$ signals, and a test set with $1,704$ signals. The three datasets were divided randomly, ensuring distributions that are statistically similar, with the only difference being the restrictions on the parameter ranges. The larger parameter ranges on the training and validation set make the emulator capable of exploring a variety of astrophysical scenarios, but we keep narrower ranges on the parameters in the test set to make the results comparable to \textsc{21cmGEM} and \textsc{globalemu}.

Before training the emulator, the training signals are preprocessed. This is done by computing the mean temperature at each frequency across all signals and subtracting this from the respective signals. Afterwards, every signal is divided by the standard deviation across all signals and frequencies. The preprocessed signals thus have zero mean in every frequency bin and are scaled to units of standard deviation. Standard practice is to divide by the standard deviation in each frequency bin; we deviate from this because the low standard deviation between signals at small frequencies would make the preprocessed signals blow up.

We train the emulator on the signals in the training set, while monitoring the performance on the training set and validation set. Only the errors on the training set are being back-propagated through the neural networks; thus, the errors on the validation set measures whether the model is able to emulate signals that are statistically similar to the training signals but not propagated through the network. The validation error is therefore an important metric for overfitting and the ability of the emulator to generalize to unseen signals. We use it to regulate the learning rate of the model, to trigger early stopping, and during hyperparameter tuning: see sections \ref{subsec:training} and \ref{subsec:tuning} for more details. 

The test set is not seen by the emulator until after training and hyperparameter tuning. The performance on the test set is therefore a measure of how well the emulator can predict global signals from parameters it was not optimized for. Hence, the test error is the best estimate of the performance of the emulator in a real setting and is therefore the figure of merit we use.

\subsection{Training} \label{subsec:training}
We trained the emulator for 350 epochs, using minibatches with 256 signals in each, and a loss function defined as the square of the figure of merit (Eq. \ref{eq:fom}). We used the \textit{Adam} \citep{2014arXiv1412.6980K} optimizer and an initial learning rate of 0.01. Together, these define the gradient descent algorithm used to minimize the loss function.

During training, we also used a learning rate schedule, which reduces the learning rate if the validation loss does not decrease over 5 epochs, and early stopping which stops the training if the validation loss does not decrease over 15 epochs. We also saved the model weights and biases after each epoch and loaded the weights and biases that correspond to the epoch with the smallest validation loss after training. This is normally towards the end of the training, but will be at an earlier epoch in the case of overfitting.

\subsection{Hyperparameter Tuning} \label{subsec:tuning}
In our neural networks, there are several parameters that cannot be optimized during training. These are called hyperparameters; examples include the number of layers in the network, dimensionality of each layer, and the loss functions. We performed hyperparameter tuning to optimize the number of layers and dimensionality of each layer. Although there exist dedicated Python packages to do this, we wrote our own script based on a random grid search to get full control over the tuning. We specified an appropriate hyperparameter space, used the script to randomly generate parameters in the space and built the emulator with these parameters. The model was trained and the validation loss of the emulator was saved. We trained 500 emulators with different hyperparameters. The hyperparameters associated with the emulator with the lowest validation loss were used to build the final product; these define the architecture which is described in section \ref{subsec:architecture}.

\section{Performance}
\label{sec:results}
\subsection{Test Errors}
\begin{figure}
    \centering
    \includegraphics[width=\linewidth]{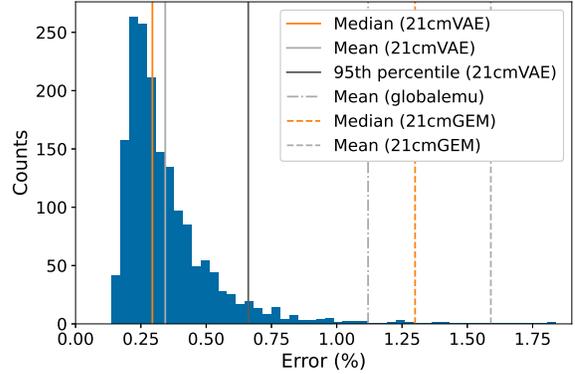}
    \caption{The histogram shows the distribution of errors for the test set of 1704 astrophysical signals with mean, median, and 95th percentile marked (solid lines). It also shows the results for \textsc{globalemu} (dash-dotted line) and \textsc{21cmGEM} (dashed lines). Note that \textsc{globalemu} does not report the median error and that the 95th percentile errors of  \textsc{globalemu} and \textsc{21cmGEM} are 2.41\% and 3.49\% respectively, hence out of the range used for the x-axis here.}
    \label{fig:hist}
\end{figure}

The training algorithm is stochastic, hence training the same model twice on the same dataset may yield different results. We therefore trained the emulator 20 times after optimizing the architecture with hyperparameter tuning. For each of the 20 trials, we computed the error for each signal in the test set. We get a mean error across signals for each trial; the distribution of these mean errors is showed in Appendix \ref{appendix:21cmAE}, in Figure \ref{fig:error_distribution}. {\textsc{21cmVAE}} has a mean error across the 20 trials of $0.354\% \pm 0.001\%$ and a median error of $0.305\% \pm 0.001\%$. The results of the best trial of {\textsc{21cmVAE}} are displayed in the histogram in Figure \ref{fig:hist}, which also shows how the performance compares to \textsc{21cmGEM} and \textsc{globalemu}.

We list the relative (as defined in Eq. \ref{eq:fom}) and absolute (the rms in mK) errors of the best trial across the entire simulated frequency range (approximately 28-237 MHz) and for selected frequency bands in Table \ref{tab:results}:
50-100 MHz (corresponding to EDGES Low Band), 60-120 MHz (corresponding to EDGES Mid Band), 90-200 MHz (corresponding to EDGES High Band), and 50-200 MHz.
\begin{table*}
    \centering
    \begin{tabular}{c c c c c c}
    \hline
       $\nu$ (MHz) & $z$  & Mean Error (\%) & Median Error (\%) & Mean Error (mK) & Median Error (mK)  \\
         \hline
         28-237 & 5.0-50.0 & 0.34 & 0.29 & 0.54 & 0.50 \\
         \hline
         50-100 & 13.3-27.4 & 0.36 & 0.29 & 0.50 & 0.45 \\
         \hline
         60-120 & 10.9-22.6 & 0.38 & 0.30 & 0.57 & 0.52 \\
         \hline
         90-200 & 6.2-14.7 & 0.82 & 0.56 & 0.90 & 0.79 \\
         \hline
         50-200 & 6.2-27.4 & 0.45 & 0.39 & 0.71 & 0.65\\
         \hline
    \end{tabular}
    \caption{The mean and median of the error of {\textsc{21cmVAE}} across the full frequency range the global signals are sampled at, and across examples of frequency bands used by the EDGES experiment.
    \label{tab:results}}
\end{table*}
The mean error across the entire frequency range is 0.54 mK or 0.34\% of the signal amplitude. For comparison, the rms between the global signal predicted from all parameters in the middle of the test range and the global signals of a model with all the parameters simultaneously shifted by $\pm 1\%$ is 0.70 mK. Hence, the emulator is sensitive to $\mathcal{O}(1\%)$ changes in the astrophysical input parameters.

\begin{figure*}[!ht]
    \centering
    \includegraphics[width=\textwidth]{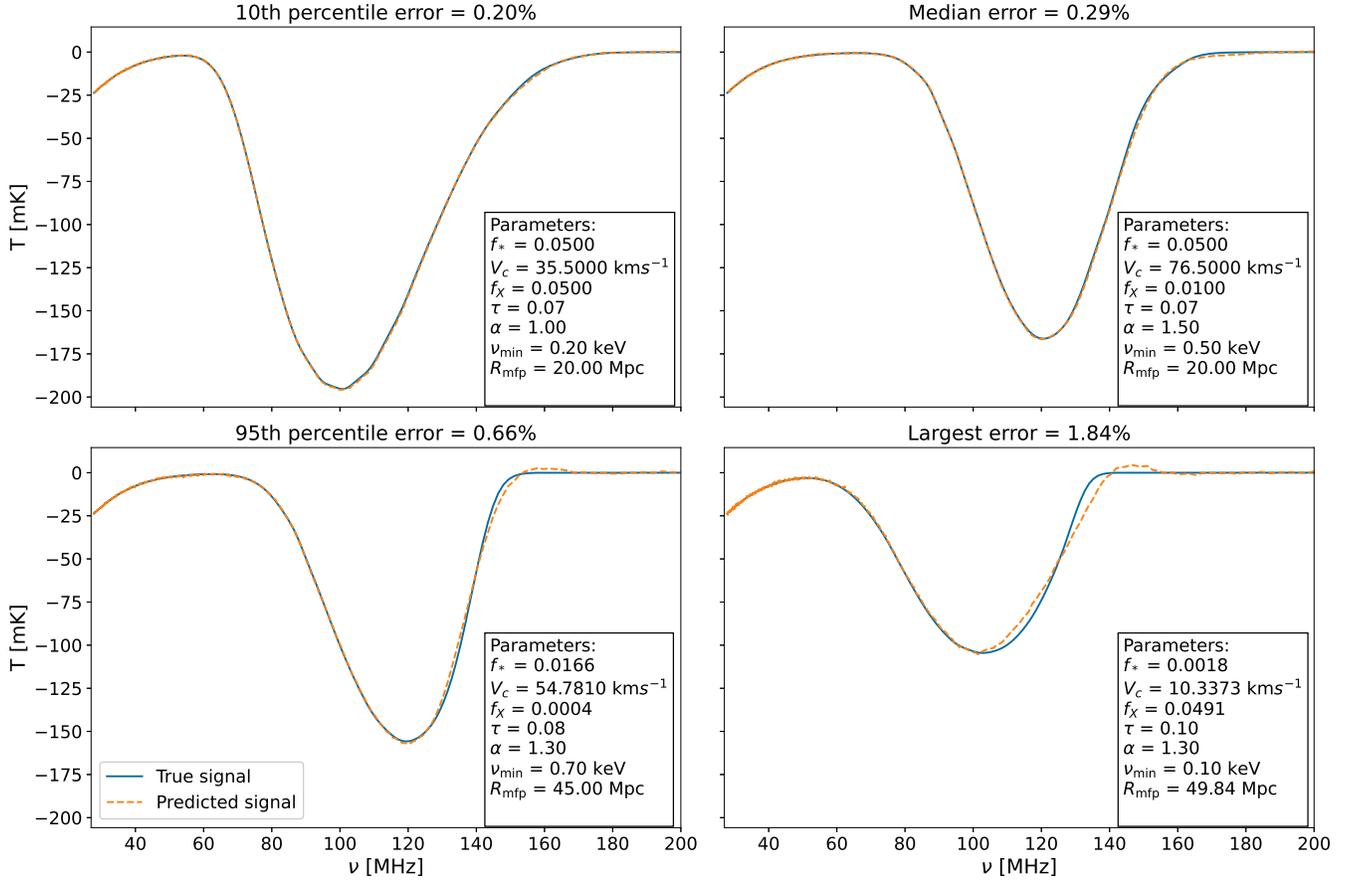}
    \caption{Comparison of true signal (blue solid line) in the test set and emulated signals (orange dashed line) for the case with error closest to the 10th percentile, median, 95th percentile, and the largest error. The errors and the values of the astrophysical parameters of each corresponding model  are shown in each panel of the figure.}
    \label{fig:comparison}
\end{figure*}

For a sample view of the performance of {\textsc{21cmVAE}} on the test set, we show in Figure \ref{fig:comparison} the predicted signals with error closest to the 10th percentile, the median, and the 95th percentile, as well as the signal with the largest error. This can be directly compared to the analogous Figure 9 of \citet{2021MNRAS.508.2923B} and Figure 13 of \citet{2020MNRAS.495.4845C}, which show the performances of \textsc{globalemu} and {\textsc{21cmGEM}}, respectively.

\subsection{Speed}
We measure the time between inputting astrophysical parameters to when an output is produced on a computer with a GeForce RTX 2080 Ti Rev. A. GPU. By randomly drawing parameters from the parameter ranges of the training set a thousand times, we find that it takes on average $0.0414 \pm 0.0007$ s to predict one global signal. We also measured the time it takes to predict 1000 signals simultaneously---that is, inputting 1000 parameter combinations to the emulator at once---obtaining an average time of $0.0418 \pm 0.0002$ s.

\section{Impact of Astrophysical Parameters}
\label{sec:impact}
We use {\textsc{21cmVAE}} to investigate the impact of each astrophysical parameter on the global signal, both qualitatively and quantitatively. First, we show visually how each parameter affects the emulated global signal. We use the signal emulated from the mean parameters in the test set as our nominal model and vary each parameter uniformly between its minimal and maximal value in the test set. The mean parameters are given by $\bar X \equiv \{\log{f_*}= -0.800$, $\log{(V_c/1 \textrm{km s}^{-1})}= 1.4803$, $\log{f_X}= -0.1190$, $\tau= 0.07$, $\alpha= 1.25$, $ \nu_{\small \textrm{min}}= 0.85$ keV, $R_{\textrm{\small mfp}}= 29.60$ Mpc\}.

\begin{figure*}
    \centering
    \includegraphics[width=\textwidth]{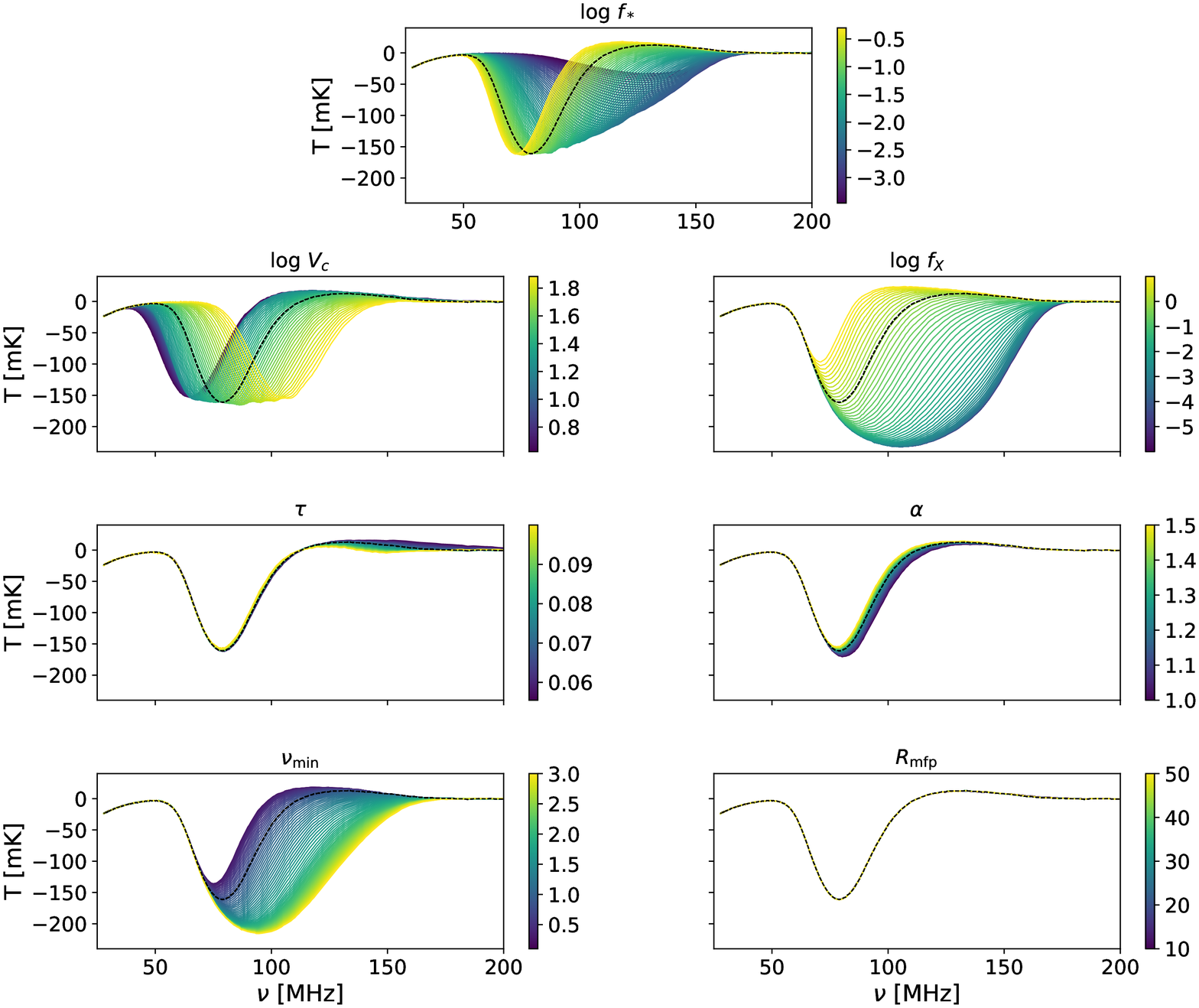}
    \caption{The effect of each astrophysical parameter on the emulated global signal. The black, dashed line in each panel is the nominal model. Each panel shows the global signal obtained by varying one parameter, while keeping all the others fixed at their nominal values. The color of each signal represents the value of the varied parameter---on a logarithmic (base 10) scale in the cases of $f_*$, $V_c$, and $f_X$.}
    \label{fig:vary_params}
\end{figure*}

Figure \ref{fig:vary_params} shows the effect of each parameter on the global signal. In general, larger values of $f_*$ and smaller values of $V_c$ are both associated with the center of the absorption trough being shifted to earlier times (lower frequencies); this is as expected since this combination of parameters corresponds to earlier star formation. Furthermore, we note that the amplitude of the global signal is regulated by X-ray heating: it increases with decreasing $f_X$ and increasing $ \nu_{\small \textrm{min}}$, both of which correspond to less X-ray heating. Less heating also lead to later absorption, as can be seen in the figure. Whereas the four parameters discussed have combined a significant impact on the center, amplitude, and width of the absorption trough, the global signal is much less sensitive to the last three parameters. As anticipated, $\tau$ is only important at late times and regulates the amplitude during reionization. The variations in amplitude are below the 10 mK-level, but the 21-cm line can be seen in emission at low redshifts for sufficiently small values of $\tau$. Being associated with X-ray heating, the global signal varies with $\alpha$ along the same trends as it varies with $f_X$ and $ \nu_{\small \textrm{min}}$, but the variations are smaller and $\alpha$ is more constrained than the other parameters. Finally, $R_{\textrm{\small mfp}}$ has no apparent effect on the global signal at the scales considered here.

Since {\textsc{21cmVAE}} maps astrophysical parameters to realizations of the global signal, we can compute the derivative of this map with respect to the input parameters. This allows us to quantify the impact \textit{I} of each astrophysical parameter on the global signal, which we define as the rms of the derivative of the mean global signal---that is, the global signal corresponding to the mean parameters of the test set ($\bar X$)---with respect to its inputs. We define the impact of each parameter $\bar X_i$ by
\begin{equation}
    I(\bar X_i) \equiv \sqrt{{\Bigg \langle \left( \frac{\rm{d}\bar T(\nu)}{\rm{d}X_i} \right)^2 \bigg \rangle}_{\nu}}\Delta X_i.
\end{equation}

We also compute the impact of the combinations $\log{(f_* \cdot f_X)}$, which is proportional (up to the minor effect of X-ray SED) to the total amount of X-rays that goes into heating of the IGM, and $\log{(f_*/V_c^3)}$, which measures the efficiency of the Wouthuysen-Field effect and is proportional to the total mass of gas in dark matter halos converted into stars \citep[see e.g][]{2016PhR...645....1B}. The first factor in the above equation, $\sqrt{{\Bigg \langle \left( \frac{\rm{d}\bar T(\nu)}{\rm{d}X_i} \right)^2 \bigg \rangle}_{\nu}}$, is proportional to the Fisher information for an instrument that has constant error bars for each pixel in the given frequency band. The other factor, $\Delta X_i$, is the uncertainty on the parameter, which we take to be the width of the parameter range in the test set. Table \ref{tab:derivs} show the impact of each parameter in the frequency bands that we reported the results for in Table \ref{tab:results}. While this analysis considers derivatives at the mean parameters in our test set, it could be repeated at any point of interest in the parameter space.

\begin{table*}
    \centering
    \begin{tabular}{c c c c c c c c c c c}
    \hline
    $\nu$ (MHz) & z & I($\log{f_*}$) & I($\log{V_c}$) & I($\log{f_X}$) & I($\tau$) & I($\alpha$) & I($\nu_{\small \textrm{min}}$) & I($R_{\textrm{\small mfp}}$) & I($\log{(f_* \cdot f_X)}$) & I($\log{(f_*/V_c^3)}$) \\
         \hline
         28-237 & 5.0-50.0 & 64.56 & 91.64 & 121.46 & 6.42 & 13.88 & 114.28 & 0.49 & 269.28 & 218.68 \\
         \hline
         50-100 & 13.3-27.4 & 104.76 & 156.36 & 189.19 & 6.71 & 21.57 & 177.64 & 0.79 & 420.62 & 365.40 \\
         \hline
         60-120 & 10.9-22.6 & 124.60 & 175.36 & 236.13 & 8.27 & 26.99 & 222.06 & 0.85 & 521.96 & 420.31 \\
         \hline
         90-200 & 6.2-14.7 & 98.47 & 107.93 & 224.83 & 13.09 & 25.77 & 211.76 & 0.47 & 486.01 & 293.91 \\
         \hline
         50-200 & 6.2-27.4 & 93.88 & 133.26 & 176.63 & 9.27 & 20.19 & 166.18 & 0.69 & 391.53 & 318.00 \\
         \hline
    \end{tabular}
    \caption{The impact in mK of each parameter on the mean global signal across the frequency bands considered in Table \ref{tab:results}.}
    \label{tab:derivs}
\end{table*}

The results in Table \ref{tab:derivs} indicate that the global signal at redshifts $z=5-50$ is most sensitive to changes in the combinations $f_* \cdot f_X$ and $f_*/V_c^3$. The four parameters with the greatest impact in this redshift range are $f_X$, $\nu_{\small \textrm{min}}$, $V_c$, and $f_*$ and the three parameters with the smallest impact are $R_{\textrm{\small mfp}}$, $\tau$, and $\alpha$. The four parameters with the largest quantitative impact are the same as the parameters that visually appeared to have the most impact on the global signals. Thus, the quantitative impact analysis confirms the intuition provided in Figure \ref{fig:vary_params}: The global signal is most sensitive to the strength and time of onset of star formation and heating in the redshift range $z=5-50$.

Table \ref{tab:derivs} also shows how the impact of the parameters changes with the observed frequency band, hence how it changes with time. All the parameters except for $\tau$ have their largest impact in the redshift range $z=10.9-22.6$, which is expected since the global signal used in this analysis both have its absorption through and go into emission in this redshift range. The most important parameters in this range are $\log{(f_*\cdot f_X)}$ and $\log{(f_*/V_c^3)}$ with impact values of 521.96 mK and 420.31 mK, respectively. These also have the greatest impact in the range $z=6.2-14.7$, but their impact decrease to about 90\% and 70\% respectively of their values in the range $z=10.9-22.6$. The only parameter that increases in impact is $\tau$, which goes from 8.27 mK to 13.09 mK. This shows that $\tau$ becomes comparatively more important at lower redshifts. On the other hand, the other reionization parameter, $R_{\textrm{\small mfp}}$, has consistently the smallest impact and does not appear to become more important at later times.

\section{Interpreting the Latent Space}
\label{sec:tsne}
Having seen how the global signals vary with the astrophysical parameters, we analyze how the last hidden layer change with the parameters. We denote this representation the latent representation or the latent space. It is 224-dimensional, since the last hidden layer has 224 neurons (in comparison, the frequency representation is 451-dimensional since the global signals are sampled at 451 frequencies). The latent representation is related to the output global signals by only one linear transformation and the additive bias of the neural network. Investigating the structure of this representation can therefore illustrate which of the seven astrophysical parameters have the strongest effect on the global signals in the frequency range $\nu=28-237$ MHz. 

\begin{figure*}
    \centering
    \includegraphics[width=\textwidth]{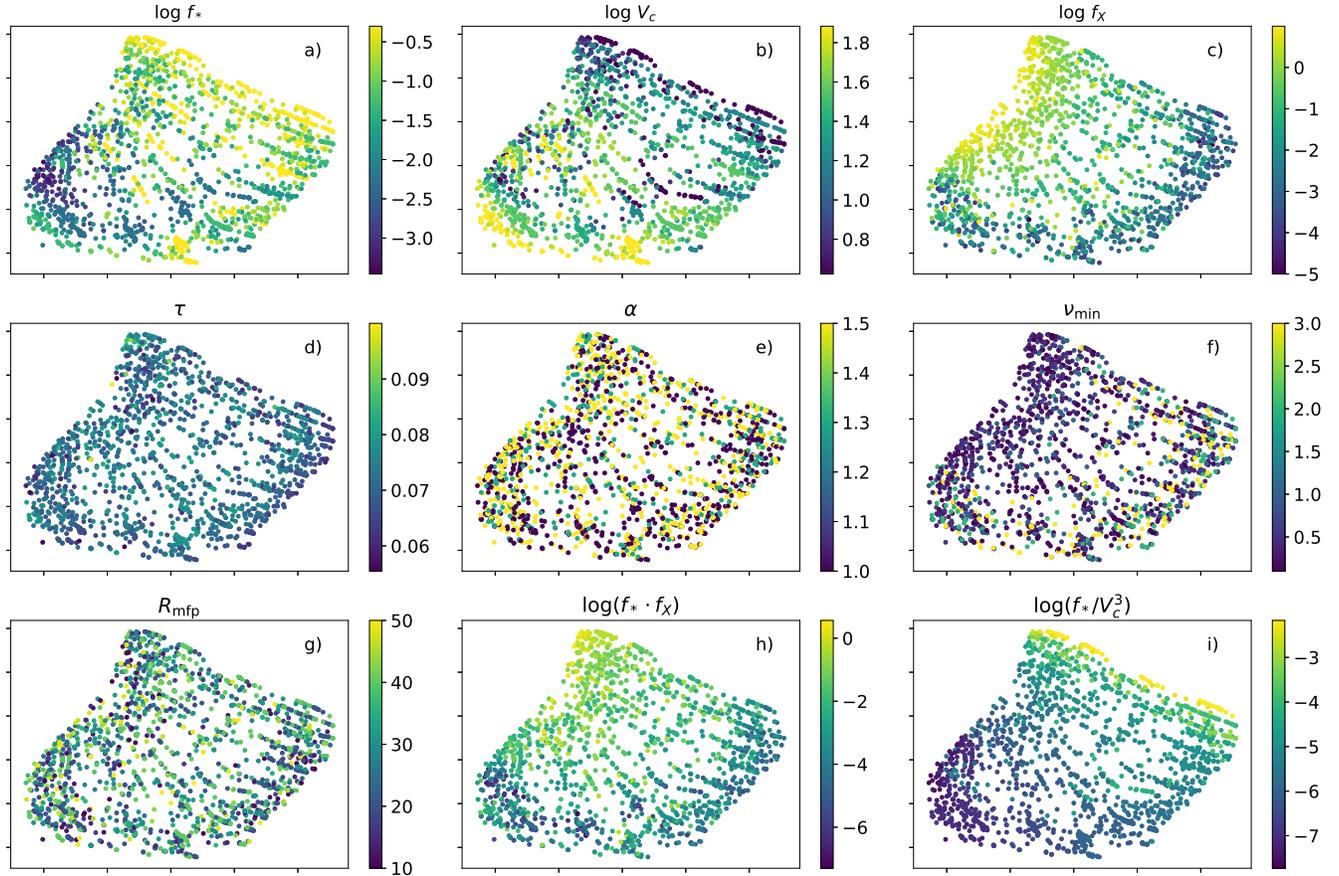}
    \caption{A t-SNE visualization of the latent representation colored by the seven astrophysical input parameters (panels a)-g)) and the combinations $f_* \cdot f_X$ (panel h)) and $f_*/V_c^3$ (panel i)). The points shown are the values of the neurons in the last hidden layer of the emulator associated with each global signal. The first three parameters and the two combinations are given on log scales (base 10), otherwise the units are the same as used elsewhere in the text. In general, the parameters that are most important for the global signal in the frequency band $\nu=28-237$ MHz are most separated in the t-SNE representation of the latent space.}
    \label{fig:tsne}
\end{figure*}

As an illustration, we plot (Figure \ref{fig:tsne}) the value of each of the 224 neurons in the last hidden layer for each of the $1,704$ signals from the test set projected to a plane using the t-SNE technique \citep[t-Stochastic Neighbor Embedding, ][]{JMLR:v9:vandermaaten08a}. This technique preserves relative distances, meaning that points close in the latent space are also close in the 2D projection. We require that the 200 closest points, or about 10\%, are counted as neighbors (i.e., setting the perplexity parameter for t-SNE to 200). For each of the seven astrophysical parameters, we color a copy of this plot according to the value of the parameters and show the results in Figure \ref{fig:tsne}. To gain further insight we also show the combinations $f_* \cdot f_X$ and $f_*/V_c^3$ discussed in Section \ref{sec:impact}. 

Since Figure \ref{fig:tsne} is an attempt at visualizing a 224-dimensional space in 2D, the shapes and structures are not very meaningful. However, as t-SNE preserves relative distances, the relative placement of the points contains information. By visually exploring Figure \ref{fig:tsne}, we find, as anticipated, that the structure of the latent space is most sensitive to changes in the astrophysical parameters regulating star formation (namely $f_*$ and $V_c$ shown in panels a and b respectively) and X-ray heating (via the dependence on $f_X$ and, to a lesser degree, on $\nu_{\small \textrm{min}}$, panels c and f respectively) as these parameters have the sharpest effect on the global signal within the explored frequency range. We can also see how the latent space is structured: The upper right part of the space is a region with large $f_*$ (panel a) and small $V_c$ (panel b), indicating that this part of the latent space is associated with early star formation. This behavior is even more clearly seen on panel (i) which is colorcoded with respect to $f_*/V_c^3$. Similarly, the latent space separates $f_X$ (panel c), with large values to the upper left and smaller values to the lower right and $\nu_{\textrm{\small min}}$ (panel f) with smaller values to the upper left and larger values to the lower right, effectively corresponding to stronger and earlier heating respectively \citep{2014Natur.506..197F} and, thus, agreeing with the distribution of $f_X$. In other words, the amount of heating generally increases upwards and to the left in the t-SNE representation of the latent space, which is highlighted in panel (h) corresponding to $f_* \cdot f_X$. 

We also find that the effects of the parameters regulating reionization (in our case it is the CMB optical depth and the mean free path of the ionizing photons, panels d and g respectively) as well as the slope of X-ray SED (panel e) is not reflected in the t-SNE plots and there is no apparent structure across the corresponding panels. In order to minimize the loss of the emulator, it was not necessary to organize the hidden layer in a way that sorts these parameters in the t-SNE projection, indicating that they have smaller effects on the global signals when explored across the broad frequency range $\nu \approx 28-237$ MHz. 

In summary, we find---as expected---that processes of X-ray heating and Wouthuysen-Field coupling have the strongest impact on the global signal in the explored frequency range with the latent space separating the amount of star formation along one diagonal in the t-SNE representation and the X-ray heating along the other diagonal. In contrast, the parameters regulating reionization as well as the slope of X-ray SED have a subdominant effect on the apparent structure of the latent space within the broad frequency range considered.

This is largely consistent with our findings in section \ref{sec:impact}, where we saw visually that the center, amplitude, and width of the global signals are most sensitive to the amount of X-ray heating and star formation, and quantitatively that the parameters regulating these effects have the greatest impacts. Larger amplitudes are associated with less X-ray heating and, to a lesser degree, more star formation. Since these are largely separated along different diagonals, the t-SNE representation does not separate amplitudes to the same extent as it separates the amount of X-rays and star formation. However, we may loosely identify the lower, right region of the latent space with global signals of large amplitude, since it is a region with low $f_X$ and, in general, above average values of $f_*$ and $ \nu_{\small \textrm{min}}$. The opposite side of the plot is a region of smaller amplitudes, with large $f_X$, more small values of $f_*$ and small $ \nu_{\small \textrm{min}}$. On the other hand, the center and width of the global signals appear to be separated with increasing width and the center shifted to high frequencies in the bottom half of the latent space (following generally the separation of $f_*\cdot f_X$).

The parameters $\tau$, $\alpha$, and $R_{\textrm{\small mfp}}$ both have the smallest effect on the global signal and do not separate the latent space in the t-SNE representation. This shows that the emulator, as anticipated, does not prioritize separating parameters that do not significantly change the shape of the global signal.

\section{Conclusions} \label{sec:conclusions}
We have presented {\textsc{21cmVAE}}, a new emulator of the 21-cm global signal, which predicts signals based on a seven-parameter input. The prediction pipeline is simple, with one neural network, whose architecture is optimized with hyperparameter tuning. The code used to optimize the network is publicly available, together with the final product. {\textsc{21cmVAE}} emulates global signals in approximately 0.04 seconds on average and with a mean rms error of 0.34\% of the signal amplitude, a significant improvement in error from existing emulators. The absolute error is on average smaller than the rms between two signals with parameters changed by 1\%.

Compared to \textsc{globalemu}, which is the fastest and most accurate existing emulator of the global 21-cm signal, as well as the most similar in architecture, \textsc{21cmVAE} has overall smaller errors in prediction of the same global signals: both the mean and maximum errors of \textsc{21cmVAE} are more than a factor of 3 smaller than the mean and maximum errors of \textsc{globalemu} over the redshift range $z=5-50$. This is due to different priorities: \textsc{globalemu} aims to be as simple and fast as possible given a target accuracy of about 10\% of the expected noise of the REACH experiment, whereas \textsc{21cmVAE} aims to be as accurate as possible without compromises. \textsc{globalemu} is indeed able to emulate the global signals in 1.3 ms, which is even faster than \textsc{21cmVAE}.

By exploring different representation of the global signal and its derivatives, we were able to qualitatively and quantitatively establish which of the model parameters create the most significant changes in the global 21-cm signal across a broad frequency range. As anticipated, we find that processes of X-ray heating and Wouthuysen-Field coupling have the strongest impact, while the parameters regulating reionization as well as the slope of X-ray SED have no apparent effect on the latent representation for the redshift range $z=5-50$. As the analysis in Section \ref{sec:impact} show, the impact of each parameter depends on the frequency band considered, and the latent representation will therefore depend on the frequency sampling of the global signals in the training set. 
The visual latent representation and the quantitative derivative analysis are a potentially powerful diagnostic that can point out dominant astrophysical processes and help optimizing theoretical modeling when targeting specific frequency bands of different experiments.

In summary, \textsc{21cmVAE} achieves unprecedentedly small errors for a range of 21-cm models across a wide frequency band. Combined with the short running time and the implementation in Python, this makes {\textsc{21cmVAE}} ideal for parameter fitting such as MCMC.

\section*{Acknowledgments}
SKNP acknowledges support from the DIRAC Institute in the Department of Astronomy at the University of Washington. The DIRAC Institute is supported through generous gifts from the Charles and Lisa Simonyi Fund for Arts and Sciences, and the Washington Research Foundation. AF was supported by the Royal Society University Research Fellowship.

\appendix

\section{Autoencoder-based Emulator} \label{appendix:21cmAE}

In addition to {\textsc{21cmVAE}}, we trained an autoencoder to generate a low-dimensional---latent---representation of the global signals and an emulator based on the autoencoder. These two networks were trained the same way as {\textsc{21cmVAE}} and the hyperparameters were optimized with the same hyperparameter tuner. The emulator maps astrophysical parameters to the latent representation, which we set to 9 dimensions after hyperparameter tuning, and then decoded by the autoencoder. By compressing the data to the latent space, the autoencoder is forced to learn the most robust features of the global signal. We believed that this would aid the emulator in predicting the global signals and that the 9 latent parameters would be easier to interpret as combinations of the input astrophysical parameters than the hidden layers of {\textsc{21cmVAE}} which all have more than 200 dimensions. Despite this, {\textsc{21cmVAE}} actually performs better than the autoencoder-based emulator and can be interpreted both qualitatively and quantitatively, as done in sections \ref{sec:impact} and \ref{sec:tsne}. We still show the methods and results here for comparison.

The autoencoder attempts to reconstruct global signals but has a 9-dimensional latent layer which forces it to reduce the dimensionality of each global signal from 451 to 9. The encoder has one hidden layer with 352 dimensions, whereas the decoder has two hidden layers with 32 and 352 dimensions, respectively. The emulator takes in the 7 astrophysical parameters and outputs the latent representation, using four hidden layers of dimensions 352, 352, 352, and 224.

\begin{figure}[!htb]
    \centering
    \includegraphics[width=\linewidth]{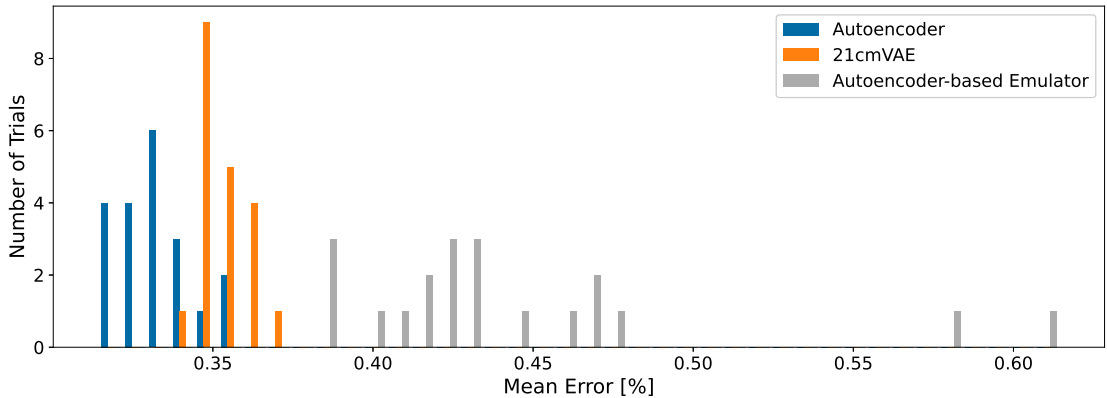}
    \caption{Histogram of the mean errors of the 20 trials for the autoencoder (blue), {\textsc{21cmVAE}} (orange), and the autoencoder-based emulator (gray).}
    \label{fig:error_distribution}
\end{figure}

As with {\textsc{21cmVAE}}, we trained the tuned autoencoder-based emulator 20 times and computed the mean test error for each trial. These are shown in Figure \ref{fig:error_distribution}.

{\textsc{21cmVAE}} has a mean error across trials of $0.354\% \pm 0.001\%$ and a median error of $0.305\% \pm 0.001\%$, whereas the autoencoder-based emulator has a mean error of $0.44\% \pm 0.01\%$ and a median error of $0.39\% \pm 0.01\%$. We see that the distribution of mean errors of {\textsc{21cmVAE}} is strictly at smaller errors than the distribution of errors of the autoencoder-based emulator. However, the errors due to the autoencoder itself are in general smaller than the errors of {\textsc{21cmVAE}}, as shown in Figure \ref{fig:error_distribution}. The autoencoder has a mean error of $0.332\% \pm 0.003\%$ and a median error of $0.292\% \pm 0.002\%$. The errors of autoencoder-based emulator would be identical to the autoencoder errors if it could perfectly map the astrophysical parameters to the autoencoder latent parameters; thus, the difference between these distributions is due to errors in the map from astrophysical parameters to latent parameters. An improved autoencoder-based emulator could reduce this difference, but the autoencoder errors are likely the limit of how small the autoencoder-based emulator errors can be. For the best trial, the mean error of the autoencoder-based emulator on the test set is $0.39\%$ and the median error is $0.35\%$. This is worse than {\textsc{21cmVAE}} (mean: $0.34\%$, median: $0.29\%$), but a significant improvement over {\textsc{21cmGEM}} (mean: $1.59\%$, median: $1.30\%$). The autoencoder that is used by that emulator has a mean error of $0.33\%$ and a median error of $0.29\%$.

\section{Dataset size} \label{appendix:dataset}
\begin{figure}[!h]
    \centering
    \includegraphics[width=\linewidth]{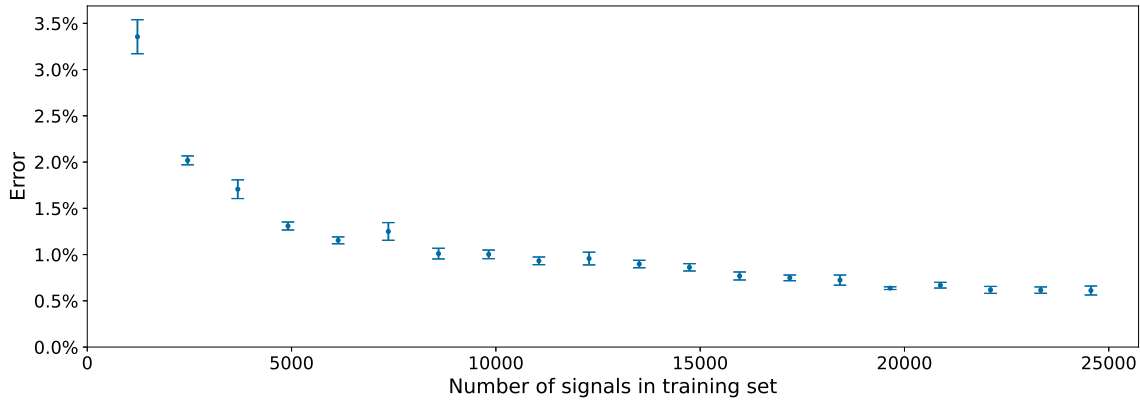}
    \caption{Test error versus the size of the training set as a percentage of the full training set for the emulator. The points and errorbars represent the mean and standard error in the mean respectively.}
    \label{fig:dscomparison}
\end{figure}

The performance of the emulator is correlated with the size of the training dataset. To test this, we randomly sample subsets of the training set and evaluate the performance of the emulator trained on each of the subsets. The subsets range in size from 5\% to the full dataset in increments of 5\%. We sample the subsets at each given size ten times and average the error across the samples. This ensures that the results only depends on the size of the subsets and not the composition of the subsets. This relationship is displayed in Figure \ref{fig:dscomparison}. We see that the emulator error steadily decreases with increasingly more signals until the dataset size is $\sim 35\%$ (about 8600 signals), after which it only slowly decrease. Therefore, the marginal effect of increasing the size of the training set used to train {\textsc{21cmVAE}} would likely be small.

\bibliographystyle{aasjournal}

\bibliography{VeryAccurateEmulator}

\end{document}